\documentclass[fleqn,usenatbib]{mnras}
\usepackage{newtxtext,newtxmath}
\usepackage[T1]{fontenc}
\usepackage{ae,aecompl}
\usepackage{graphicx}
\usepackage{dcolumn}
\usepackage{bm}
\usepackage{color}
\usepackage{natbib}
\usepackage{ulem}

\newcommand{\nodata}{--}
\newcommand{\KEPLER}{\textsc{Kepler}}

\title{\textsl{s}-Process in Massive Carbon-Enhanced Metal-Poor Stars}

\author[P. Banerjee et al.]{
Projjwal Banerjee,$^{1}$\thanks{E-mail: projjwal@sjtu.edu.cn}
Yong-Zhong Qian,$^{2,4}$
and Alexander Heger$^{3,4,5}$
\\
$^{1}$Department of Astronomy,  School of Physics and Astronomy, Shanghai Jiao Tong University, Shanghai 200240, China\\
$^{2}$School of Physics and Astronomy, University of Minnesota, Minneapolis, MN 55455, USA\\
$^{3}$School of Physics and Astronomy and Monash Centre for Astrophysics (MoCA), Monash University, Vic 3800, Australia\\
$^{4}$Tsung-Dao Lee Institute, Shanghai 200240, China\\
$^{5}$Monash Centre for Astrophysics (MoCA), Monash University, Vic 3800, Australia
}

\date{Accepted XXX. Received YYY; in original form ZZZ}

\pubyear{2018}
\begin{document}
\label{firstpage}
\pagerange{\pageref{firstpage}--\pageref{lastpage}}
\maketitle
\begin{abstract}
Observations suggest that the interstellar medium (ISM) might have been highly enriched in 
carbon at very early times. We explore nucleosynthesis in massive carbon-enhanced 
metal-poor (CEMP) stars of $12$--$40\,\mathrm{M}_\odot$ formed from such an ISM  with ${\rm [Fe]}\leq-2$. We find 
substantial production of elements heavier than Fe, mostly up to Sr, by the weak \textsl{s}-process in stars 
with initial abundances of ${\rm [CNO]}\gtrsim-1.5$.  Even heavier elements, up to Ba, can be produced
for ${\rm [CNO]}\gtrsim-0.5$.  The efficiency of this \textsl{s}-process is sensitive to the initial enhancement of
C (or more generally, CNO) and mass of the star, with the yield increasing approximately
linearly with the initial Fe abundance. The \textsl{s}-process in CEMP stars of $\gtrsim 20 \,\mathrm{M}_\odot$ with 
${\rm [CNO]}\gtrsim -1.5$ can be an important source for heavy elements in the Early Galaxy.
\end{abstract}


\section{INTRODUCTION}
Chemical abundances of low-mass stars typically represent the composition of the interstellar medium 
(ISM) from which they formed. Thus, surface abundances of very metal-poor (VMP) stars of 
$\lesssim 0.8 \,\mathrm{M}_\odot$ and with ${\rm [Fe/H]}=\log({\rm Fe/H})-\log({\rm Fe/H})_\odot\lesssim-2$ can provide 
a direct window on the composition of the ISM within $\sim 1\,$Gyr of the Big Bang. With a large number 
of observed VMP stars, it has now become clear that there exists a distinct class of carbon-enhanced 
metal-poor (CEMP) stars with high enhancement of C over Fe relative to the Sun. The nominal criterion 
for this classification is ${\rm [C/Fe]} \geq +0.7$ \citep{aoki2007}.  CEMP stars that do not show any 
enhancement in neutron-capture elements (${\rm [Ba/Fe]}\leq 0$), the so-called CEMP-no stars \citep{beers2005},
constitute $\sim 20\,\%$, $40\,\%$, and $80\,\%$ of all VMP stars with ${\rm [Fe/H]}\leq -2$, $-3$, and $-4$, 
are widely thought to have formed from an ISM polluted by only the first or very early core-collapse 
supernovae (CCSNe) associated with massive VMP stars.
Current models for nucleosynthesis and explosion of 
massive VMP stars can explain the abundances observed in CEMP-no stars reasonably well. These 
models involve either CCSNe of low to medium explosion energy that preferentially eject C and other 
light elements relative to the Fe group \citep{umeda2005,nomoto2006,hw2010,tominaga2014} or winds enriched 
in CNO from fast-rotating massive stars \citep{meynet2006}. In either case, the very early ISM became 
enhanced in C over Fe relative to the Sun. Low-mass stars of $\lesssim 0.8 \,\mathrm{M}_\odot$ formed from such an ISM 
would be observed as CEMP-no stars today. In contrast, massive CEMP stars of $>\!10 \,\mathrm{M}_\odot$ formed 
from the same ISM would have exploded as CCSNe within $\sim 10\,$Myr of their birth. These massive CEMP 
stars, however, can have interesting nucleosynthesis due to their C enhancement, thereby potentially 
providing an important source for chemical enrichment of the early Galaxy.

In this Letter we study the pre-CCSN nucleosynthesis of massive CEMP stars. We show that they can
produce heavy elements with mass numbers up to $A\sim 90$--$140$ by the slow neutron-capture process
(\textsl{s}-process) and serve as 
an effective source for these elements in the early Galaxy. Whereas a similar \textsl{s}-process has been 
shown to operate in fast-rotating massive VMP stars without C enhancement
\citep{pignatari2008,frischk2012,frischk2016}, our study differs in that we focus on non-rotating massive 
CEMP stars and our results are independent of the uncertain rotation-induced mixing processes.
The effects of initial composition on the \textsl{s}-process in massive stars were investigated by
\citet{prantzos1990}, \citet{baraffe1992}, and \citet{raiteri1992}. In the VMP regime of interest in this work, those
earlier studies considered the moderate enhancement of O/Fe. Whereas the effects are similar, the enhancement of 
C/Fe discussed in this work is more extreme and has a more immediate connection to the formation 
of the first stars and the earliest phase of chemical enrichment as suggested by a wide range of recent observations.
Further, the revised neutron-capture rate for $^{16}$O, the major neutron poison at low metallicities, is
$\sim 10$--100 times higher than those adopted in the earlier studies. Therefore, our work not only has new
cosmological context and implications but also provides a necessary reexamination of the earlier
results with updated nuclear physics.

\section{METHODS}
We study the nucleosynthesis in non-rotating CEMP stars of $12$--$40\,\mathrm{M}_\odot$ using 
the 1D hydrodynamical code \KEPLER{} \citep{weaver1978,rauscher2003}. 
The results from Big Bang nucleosynthesis are adopted for the initial
abundances of H to Li. Scaled solar abundances \citep{asplund2009} are assumed for all stable isotopes I
from $^9$Be to $^{70}$Zn such that ${\rm [I]}\equiv\log(X_{\rm I}/X_{{\rm I},\odot})={\rm [Fe]}$, 
except that $^{12}$C, $^{14}$N, and $^{16}$O are enhanced. Here $X_{\rm I}$ denotes the number of I atoms
per unit mass in a star. For stars with H content similar to the Sun, ${\rm [I]}\approx{\rm [I/H]}$.
Because the $^{22}$Ne providing the neutron source for the \textsl{s}-process is produced by burning 
the initial CNO, what matters is the total initial CNO abundance. Whereas we take 
${\rm [C]}={\rm [N]}={\rm [O]}={\rm [CNO]}$, the same total CNO abundance with different relative abundances 
will give identical results.  We consider ${\rm [CNO]}=-2$ to $0$ and ${\rm [Fe]}=-5$ to $-2$ but 
keeping ${\rm [C/Fe]}\geq 1$. 

We follow the nucleosynthesis from the birth of a star until its death in a CCSN using 
a large adaptive post-processing network with the same reaction rates  
as in \cite{rauscher2002}. In particular, we use the rate of \cite{jaeger2001} 
for $^{22}$Ne$(\alpha,\mathrm{n})^{25}$Mg and the rates of \cite{cf88} for 
$^{17}$O$(\alpha,\mathrm{n})^{20}$Ne and $^{17}$O$(\alpha,\gamma)^{21}$Ne.
As discussed below, these reactions are important for the \textsl{s}-process.
To include effects of the explosion on the yields of pre-CCSN nucleosynthesis,
we model the explosion by driving a piston from the base of the O shell. The velocity of the 
piston is adjusted to produce the desired explosion energy, which is taken to be $0.3\,$B, $0.6\,$B, and $1.2\,$B 
($1\,{\rm B}=10^{51}\,$ergs) for $12\,\mathrm{M}_\odot$, $15\,\mathrm{M}_\odot$, and $20$--$40 \,\mathrm{M}_\odot$ models, respectively, in order to 
match recent CCSN simulations \citep{bruenn2013,melson2015}. 
We note that explosive burning due to the CCSN shock has only minor effects on 
elements heavier than Ge. The post-CCSN yields of these elements 
are essentially the yields of the \textsl{s}-process during the pre-CCSN evolution.

\section{RESULTS}
It is well known that the weak \textsl{s}-process occurs during the pre-CCSN evolution of massive stars 
\citep{peters1968,couch1974,lamb1977,raiteri1991,pignatari2010}. The initial $^{12}$C and $^{16}$O 
in a star are first converted to $^{14}$N during core H burning. Then all of the $^{14}$N 
is converted to $^{22}$Ne via $^{14}$N$(\alpha,\gamma)^{18}$F$(e^{+}\nu_{\rm e})^{18}$O$(\alpha,\gamma)^{22}$Ne
at the start of core He burning. By the time the He core becomes convective, almost all of the initial 
CNO have been effectively converted to $^{22}$Ne, which can provide a neutron source through
$^{22}$Ne$(\alpha,\mathrm{n})^{25}$Mg during the subsequent evolution. The efficiency of the \textsl{s}-process,
however, depends on the neutron density, which is determined by the competition between production by
the neutron source and capture by neutron poisons. For a normal star,
the effects of neutron poisons render the weak \textsl{s}-process inefficient unless the initial metallicity 
of the star is ${\rm [Fe]}\gtrsim-1$. In contrast, as presented below, the enhanced initial abundances of CNO 
facilitate an efficient \textsl{s}-process in a CEMP star with ${\rm [Fe]}\lesssim-2$ and ${\rm [C/Fe]}\gtrsim 1$.

The reaction $^{22}$Ne$(\alpha,\mathrm{n})^{25}$Mg is sensitive to temperature
and is activated only when the $^4$He mass fraction 
drops to $\lesssim 0.1$ during the late stage of core He burning. The temperature
is higher for stars of higher masses. For example, when the central $^4$He mass fraction drops to 0.01,
the central temperature is $\sim 2.4\times 10^8\,$K and $\sim 2.8\times 10^8\,$K for $12 \,\mathrm{M}_\odot$ and
$25 \,\mathrm{M}_\odot$ models, respectively. This seemingly minor temperature difference corresponds to a difference 
in the $^{22}$Ne$(\alpha,\mathrm{n})^{25}$Mg rate by a factor of $\sim 50$. Consequently, in more massive stars,  
a larger fraction of $^{22}$Ne is burned to produce neutrons, which leads to a more efficient \textsl{s}-process
during the late stage of core He burning.

Regardless of its mass, the star runs out of $^4$He before $^{22}$Ne is consumed. 
In fact, most of the $^{22}$Ne survives at the end of core He burning in stars of $\lesssim 15 \,\mathrm{M}_\odot$ 
and a considerable fraction survives in those of higher masses.
The remaining $^{22}$Ne can be used for further neutron production when $\alpha$-particles are provided
through $^{12}$C$(^{12}$C$,\alpha)^{20}$Ne during the subsequent evolution. The resulting
\textsl{s}-process acts on the previous \textsl{s}-process products from core He burning. 
This additional process operates during shell C burning associated with 
core O burning and during shell He burning when the temperature at the base of the He shell increases
above $\sim 2.5 \times 10^8\,$K as the star contracts during core C and O burning.
Only a small fraction of the processed material, however, can be ejected during the CCSN, whereas most of it
becomes part of the Fe core that collapses into a neutron star or black hole. 
We find that the \textsl{s}-process mainly occurs during shell He burning for stars of $\lesssim 15 \,\mathrm{M}_\odot$ and
during core He burning for those of higher masses.

\subsection{Dependence on {\rm [CNO]}, stellar mass, and {\rm [Fe]}}

\begin{figure}
\centerline{\includegraphics[width=\columnwidth]{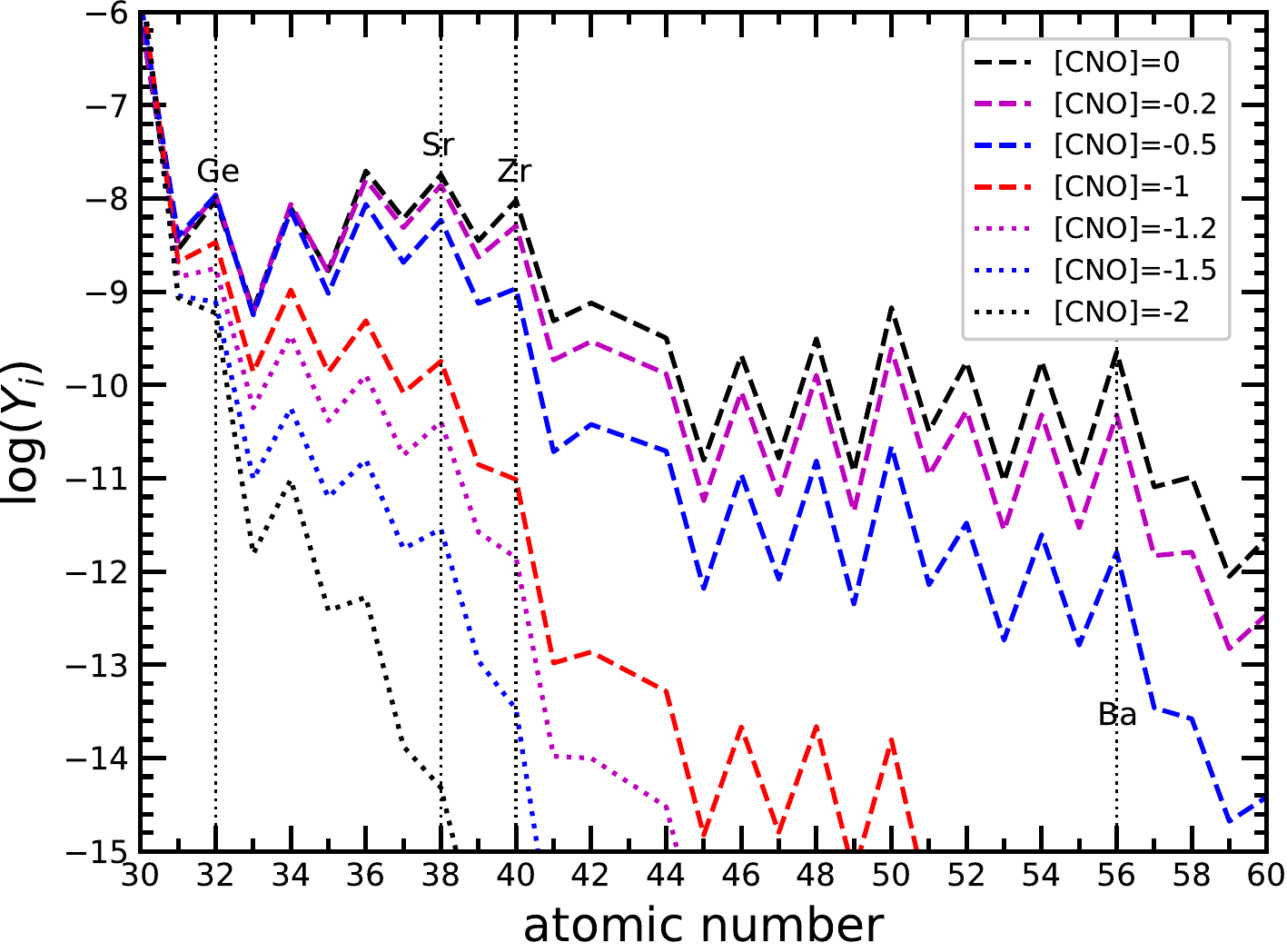}} 
\caption{Post-CCSN number yields of heavy elements for $25 \,\mathrm{M}_\odot$ models 
with a fixed ${\rm [Fe]}=-3$ but varying values of [CNO] from $-2$ to $0$.}

\label{fig:cdep}
\end{figure}

There is very little \textsl{s}-processing for ${\rm [CNO]}\leq -2$. As [CNO] increases from $-2$ to 0, the efficiency 
of the \textsl{s}-process increases dramatically (see Fig.~\ref{fig:cdep} and Table~\ref{tab:yieldmass}). 
For example, for $25 \,\mathrm{M}_\odot$ models with ${\rm [Fe]}=-3$, the Sr yield increases by a factor of 
$\sim 600$ ($4\times10^4$) when [CNO] increases from $-2$ to $-1.5$ ($-1$).  For ${\rm [CNO]}>-1$, 
the $^{22}$Ne abundance becomes so high that a significant amount of $^{25,26}$Mg is produced by 
$^{22}$Ne$(\alpha,\mathrm{n})^{25}$Mg and $^{22}$Ne$(\alpha,\gamma)^{26}$Mg. Whereas these secondary neutron 
poisons become important, their effects are more than compensated by the increased neutron production
due to the high $^{22}$Ne abundance. Consequently, the Sr yield further increases by a factor of 
$\sim 100$ when [CNO] increases from $-1$ to $0$ (see Fig.~\ref{fig:cdep} and Table~\ref{tab:yieldmass}).

\begin{figure}
\centerline{\includegraphics[width=\columnwidth]{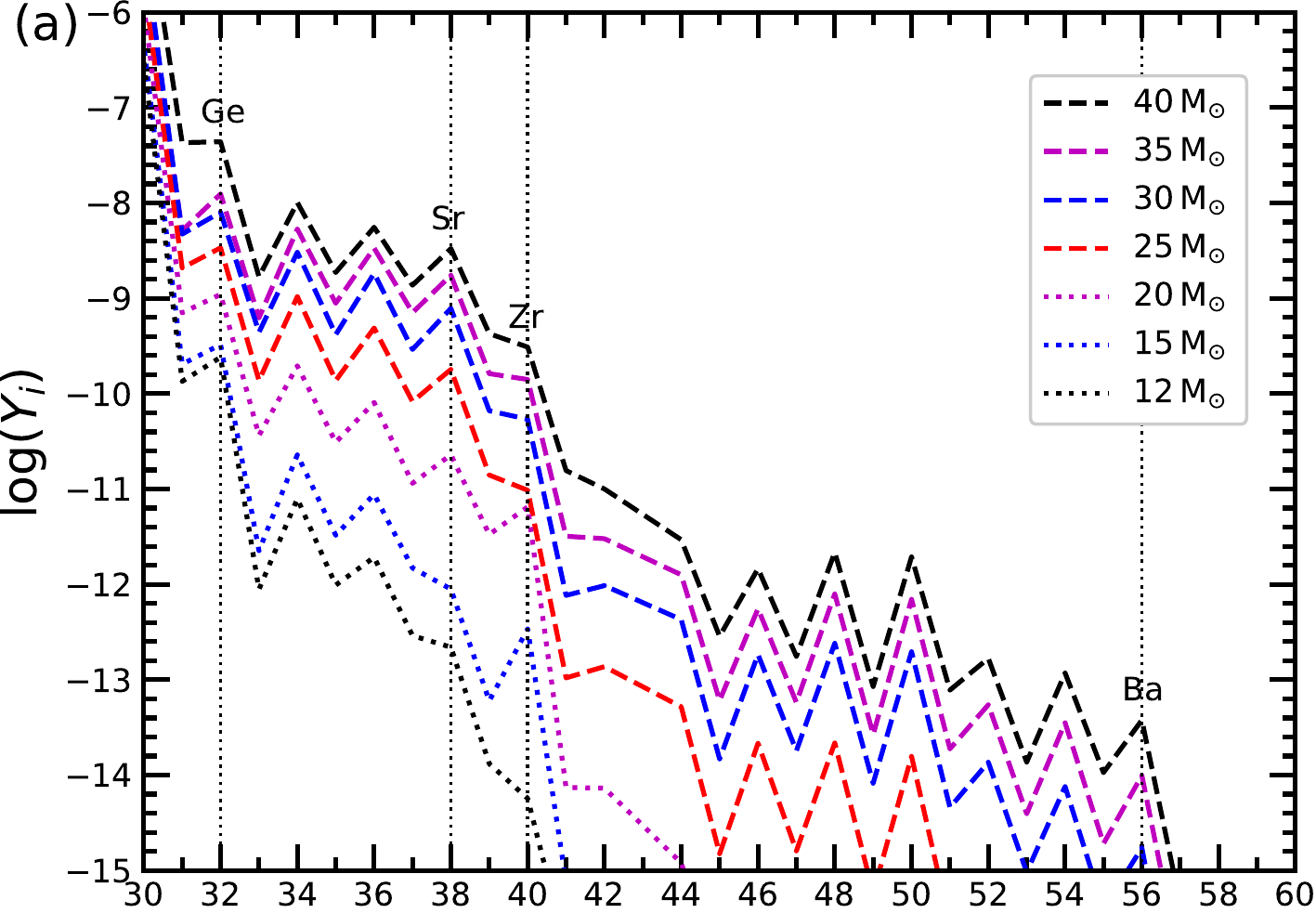}}
\centerline{\includegraphics[width=\columnwidth]{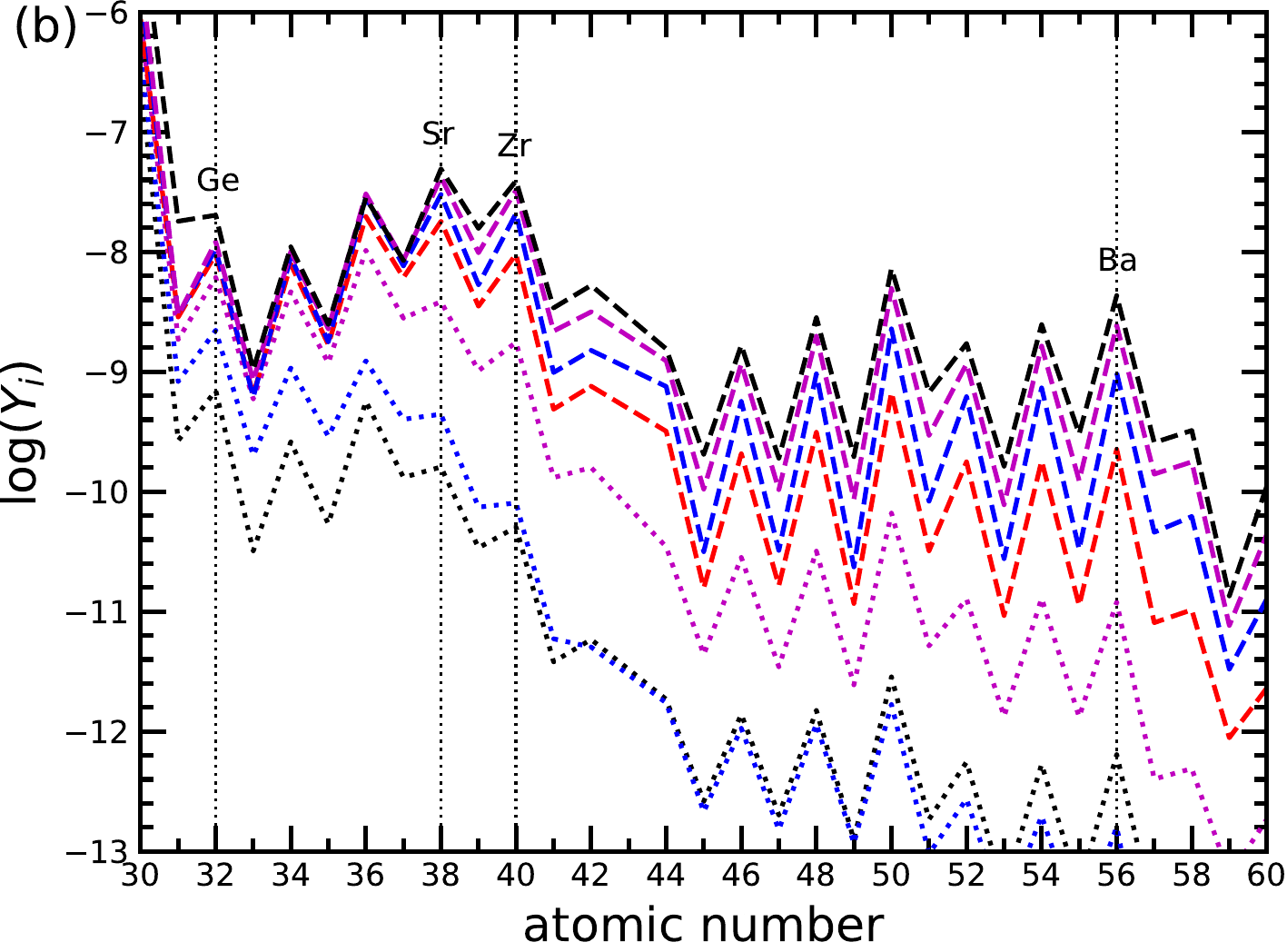}}
\caption{(a) Post-CCSN number yields of heavy elements for models with fixed ${\rm [Fe]}=-3$ and 
${\rm [CNO]}=-1$ but varying masses of 12--$40\,\mathrm{M}_\odot$. (b) Same as (a), but for models with ${\rm [CNO]}=0$.}
\label{fig:mdep}
\end{figure}

Because $^{22}$Ne$(\alpha,\mathrm{n})^{25}$Mg is very sensitive to temperature, neutron production is more 
efficient in more massive stars that burn He at higher temperature. In addition, a more massive star
has a larger He core, which allows more material to undergo \textsl{s}-processing.
The above two effects cause the efficiency of the \textsl{s}-process to increase with the stellar mass. 
For example, for models with ${\rm [Fe]}=-3$ and ${\rm [CNO]}=-1$, the Sr yield of a $12 \,\mathrm{M}_\odot$ star 
is $\sim 800$ times lower than that of a $25 \,\mathrm{M}_\odot$ star, which in turn is $\sim 10$ times lower 
than that of a $35 \,\mathrm{M}_\odot$ star (see Fig.~\ref{fig:mdep} and Table~\ref{tab:yieldmass}). 

As can be seen from the above discussion, the initial [CNO] and the mass of a CEMP star are the two 
key factors governing neutron production for its \textsl{s}-process. As discussed below, the main neutron 
poisons are the primary $^{16}$O produced by He burning and the secondary $^{25,26}$Mg produced by 
$^{22}$Ne burning. Because neither the neutron source nor the main poisons depend on the initial [Fe], 
this parameter has little impact on the efficiency of the \textsl{s}-process. Its role in this process is 
almost solely to provide the seeds for neutron capture.  Table~\ref{tab:yield25} shows the yields of 
heavy elements for $25 \,\mathrm{M}_\odot$ models with a fixed ${\rm [CNO]}=-1$ but varying [Fe].
These yields scale almost linearly with the number abundance as [Fe] increases 
from $-5$ to $-3$. The increases in the yields of Sr, Y, and Zr from ${\rm [Fe]}=-3$ to $-2$
still follow this linear scaling within a factor of $\sim2$.

\subsection{Neutron poisons}
\label{sec:poison}
Although the \textsl{s}-process in massive CEMP stars is similar to the well-known 
weak \textsl{s}-process at higher metallicities, there are some important differences,
the most crucial of which has to do with neutron poisons. As mentioned above, the \textsl{s}-process 
in CEMP stars occurs mainly during shell He burning for stars of $M\lesssim 15 \,\mathrm{M}_\odot$ and 
during the late stage of core He burning for those of higher masses. The main neutron poisons 
during these phases are the primary $^{16}$O produced by He burning and the secondary $^{25,26}$Mg 
produced by $^{22}$Ne burning. The primary nature of the former comes from the independence of the 
initial metallicity for its production, whereas the secondary nature of the latter is due to
the dependence on the initial CNO abundances for the supply of $^{22}$Ne. Because of this 
difference, $^{16}$O is the dominant poison for stars with initial abundances of 
${\rm [CNO]}\lesssim -1$ and $^{25,26}$Mg take over for ${\rm [CNO]}>-1$.

The above discussion largely holds for normal massive stars as well.
Without any CNO enhancement, however, the initial abundances of CNO for these stars are commensurate 
with those of other metals, some of which, such as $^{20}$Ne, $^{24}$Mg, and $^{28}$Si, are
also neutron poisons. The end result is that the weak \textsl{s}-process becomes efficient in normal
stars only when their initial metallicities are ${\rm [Fe]}\gtrsim -1$.
The dominant neutron poisons in this case are $^{25,26}$Mg.

\begin{table*}
\centering
\vspace{-0.5\baselineskip}
\caption{Post-CCSN yields (in $\,\mathrm{M}_\odot$) of Sr, Y, Zr, Ba, and Pb. Models are labelled as (mass/$\,\mathrm{M}_\odot$, [Fe], [CNO])
and $X(Y) \equiv X\times10^Y$. The last column gives the number of neutrons captured per available $^{56}$Fe seed,
an approximate measure of the $\textsl{s}$-process efficiency used only in \S\ref{sec:poison} for comparison with earlier studies.}
\vskip 0.1cm
\begin{tabular}{llllllrrrr}
\hline
 Model            &Sr           &Y            &Zr           &Ba          &Pb       &[Sr/Y]   &[Sr/Zr]  &[Sr/Ba] &$n_{\rm c}$\\
\hline
$(40,-3,-2.0)$      & $3.00(-11)$ &$1.09(-12)$  &$4.05(-13)$  & $0.00$        &$0.00$      &$0.74$  &$1.53$    & \nodata    &$1.69$\\
$(40,-3,-1.5)$    & $4.81(\;\;-9)$  &$3.97(-10)$  &$1.71(-10)$  & $0.00$        &$0.00$      &$0.38$  &$1.11$    &  \nodata   &$4.29$\\
$(40,-3,-1.0)$      & $2.91(\;\;-7)$  &$3.79(\;\;-8)$   &$2.80(\;\;-8)$   & $5.16(-12)$&$0.00$      &$0.18$  &$0.68$    &$4.45$  &$10.49$\\
$(40,-3,-0.5)$    & $2.89(\;\;-6)$  &$6.53(\;\;-7)$   &$9.81(\;\;-7)$   & $1.44(\;\;-8)$ &$1.35(-12)$&$-0.06$ &$0.13$  &$1.80$  &$22.64$\\
$(40,-3,\;\;0.0)$       & $4.31(\;\;-6)$  &$1.39(\;\;-6)$   &$3.61(\;\;-6)$   & $5.94(\;\;-7)$ &$1.63(\;\;-9)$&$-0.22$ &$-0.26$  &$0.36$  &$37.90$\\
 \hline
$(35,-3,-2.0)$      & $9.30(-12)$ &$2.22(-13)$  &$6.85(-14)$  & $0.00$        &$0.00$      &$0.92$  &$1.79$    &  \nodata   &$1.52$\\
$(35,-3,-1.5)$    & $2.66(\;\;-9)$  &$1.50(-10)$  &$7.42(-11)$  & $0.00$        &$0.00$      &$0.54$  &$1.21$    &  \nodata   &$3.88$\\
$(35,-3,-1.0)$      & $1.51(\;\;-7)$  &$1.43(\;\;-8)$   &$1.26(\;\;-8)$   & $1.34(-12)$&$0.00$      &$0.32$  &$0.74$    &$4.55$  &$9.29$\\
$(35,-3,-0.5)$    & $2.04(\;\;-6)$  &$3.49(\;\;-7)$   &$6.46(\;\;-7)$   & $6.21(\;\;-9)$ &$3.00(-13)$&$0.06$&$0.16$    &$2.01$  &$21.48$\\
$(35,-3,\;\;0.0)$       & $3.69(\;\;-6)$  &$8.75(\;\;-7)$   &$3.01(\;\;-6)$   & $3.35(\;\;-7)$ &$5.36(-10)$&$-0.08$&$-0.25$  &$0.54$  &$35.54$\\
\hline
$(30,-3,-2,0)$      & $2.21(-12)$ &$4.94(-14)$  &$1.50(-14)$  & $0.00$        &$0.00$      &$0.95$  &$1.83$    & \nodata    &$1.32$\\
$(30,-3,-1.5)$    & $1.01(\;\;-9)$  &$4.89(-11)$  &$1.91(-11)$  & $0.00$        &$0.00$      &$0.61$  &$1.38$    &  \nodata   &$3.31$\\
$(30,-3,-1,0)$      & $6.87(\;\;-8)$  &$5.89(\;\;-9)$   &$4.89(\;\;-9)$   & $2.35(-13)$&$0.00$      &$0.36$  &$0.81$    &$4.96$  &$8.34$\\
$(30,-3,-0.5)$    & $1.29(\;\;-6)$  &$1.83(\;\;-7)$   &$3.50(\;\;-7)$   & $2.20(\;\;-9)$&$4.67(-14)$&$0.14$ &$0.23$    &$2.26$  &$18.81$\\
$(30,-3,\;\;0.0)$       & $2.65(\;\;-6)$  &$4.73(\;\;-7)$   &$1.93(\;\;-6)$  & $1.34(\;\;-7)$ &$1.46(-10)$&$0.04$ &$-0.20$   &$0.79$  &$30.98$\\
\hline
$(25,-3,-2.0)$      & $4.17(-13)$ &$9.01(-15)$  &$3.30(-15)$  & $0.00$       &$0.00$      &$0.96$   &$1.76$    & \nodata    &$1.06$\\
$(25,-3,-1.5)$    & $2.49(-10)$ &$9.77(-12)$  &$3.01(-12)$  & $0.00$       &$0.00$      &$0.70$   &$1.58$    & \nodata    &$2.66$\\
$(25,-3,-1.2)$    & $3.53(\;\;-9)$  &$2.35(-10)$  &$1.27(-10)$  & $2.41(-18)$&$0.00$     &$0.47$   &$1.10$    &$8.66$  &$4.63$\\
$(25,-3,-1.0)$     & $1.57(\;\;-8)$  &$1.25(\;\;-9)$    &$8.79(-10)$  & $7.61(-15)$&$0.00$     &$0.40$  &$0.91$    &$5.81$   &$6.44$\\
$(25,-3,-0.5)$   & $5.15(\;\;-7)$  &$6.68(\;\;-8)$    &$9.87(\;\;-8)$   & $2.19(-10)$&$5.79(-16)$&$0.18$&$0.38$    &$2.87$  &$15.59$\\
$(25,-3,-0.2)$    & $1.21(\;\;-6)$ &$2.09(\;\;-7)$    &$4.67(\;\;-7)$   & $6.65(\;\;-9)$ &$6.40(-13)$&$0.06$&$0.08$   &$1.76$   &$22.17$\\
$(25,-3,\;\;0.0)$      & $1.58(\;\;-6)$  &$3.13(\;\;-7)$   &$8.78(\;\;-7)$   & $3.10(\;\;-8)$ &$1.28(-11)$&$0.00$  &$-0.09$    &$1.20$   &$26.41$\\
\hline
$(20,-3,-2.0)$      & $2.58(-13)$ &$2.43(-14)$  &$8.29(-14)$  & $0.00$       &$0.00$      &$0.32$  &$0.15$     & \nodata    &$0.75$\\
$(20,-3,-1.5)$    & $3.72(-11)$ &$4.19(-12)$  &$1.13(-11)$  & $0.00$       &$0.00$      &$0.24$  &$0.18$     &  \nodata   &$1.88$\\
$(20,-3,-1.0)$      & $1.99(\;\;-9)$  &$2.97(-10)$  &$5.84(-10)$  & $2.26(-17)$&$0.00$  &$0.12$  &$0.19$     &$7.44$  &$4.52$\\
$(20,-3,-0.5)$    & $8.83(\;\;-8)$  &$1.15(\;\;-8)$   &$2.21(\;\;-8)$   & $7.11(-12)$&$1.55(-17)$&$0.18$&$0.26$    &$3.59$  &$10.77$\\
$(20,-3,\;\;0.0)$      & $3.38(\;\;-7)$  &$9.07(\;\;-8)$   &$1.62(\;\;-7)$   & $1.64(\;\;-9)$ &$7.64(-13)$&$-0.13$ &$-0.02$  &$1.81$  &$18.57$\\
\hline
$(15,-3,-2.0)$      & $2.24(-14)$ &$1.31(-16)$  &$1.32(-16)$ & $0.00$       &$0.00$      &$1.53$  &$1.89$      &\nodata     &$0.45$\\
$(15,-3,-1.5)$    & $4.23(-13)$ &$9.22(-15)$  &$1.76(-13)$ & $0.00$       &$0.00$      &$0.96$  &$0.04$      &  \nodata   &$1.03$\\
$(15,-3,-1.0)$      & $7.72(-11)$ &$5.34(-12)$  &$3.07(-11)$ &$1.91(-19)$&$0.00$      &$0.46$  &$0.06$      &$8.10$  &$2.53$\\
$(15,-3,-0.5)$    & $3.63(\;\;-9)$  &$5.18(-10)$  &$9.31(-10)$ &$1.89(-13)$&$0.00$      &$0.14$  &$0.25$      &$3.78$  &$5.82$\\
$(15,-3,\;\;0.0)$      & $3.88(\;\;-8)$  &$6.60(\;\;-9)$   &$7.35(\;\;-9)$  &$2.19(-11)$&$1.59(-16)$&$0.06$&$0.38$       &$2.74$  &$11.42$\\
\hline
$(12,-3,-2.0)$      & $6.14(-15)$  &$3.43(-17)$ &$2.63(-17)$ & $0.00$       &$0.00$      &$1.55$  &$2.02$      & \nodata    &$0.33$\\
$(12,-3,-1.5)$    & $5.46(-14)$  &$7.01(-16)$ &$1.42(-16)$ & $0.00$       &$0.00$      &$1.19$  &$2.24$      & \nodata    &$0.73$\\
$(12,-3,-1.0)$      & $1.91(-11)$  &$1.17(-12)$ &$5.15(-13)$ & $0.00$       &$0.00$      &$0.51$  &$1.23$      & \nodata    &$1.76$\\
$(12,-3,-0.5)$    & $1.12(\;\;-9)$   &$1.58(-10)$ &$1.46(-10)$ &$4.10(-14)$&$0.00$      &$0.15$  &$0.55$      &$3.93$  &$4.37$\\
$(12,-3,\;\;0.0)$      & $1.41(\;\;-8)$   &$3.02(\;\;-9)$  &$4.65(\;\;-9)$  &$8.81(-11)$&$5.37(-14)$&$-0.04$ &$0.14$     &$1.70$  &$9.30$\\
\hline
\end{tabular}
\label{tab:yieldmass}
\end{table*}

\begin{table*}
\centering
\vspace{-0.5\baselineskip}
\caption{Post-CCSN yields (in $\,\mathrm{M}_\odot$) of Sr, Y, Zr, Ba, and Pb for $25 \,\mathrm{M}_\odot$ models
with ${\rm [CNO]}=-1$ but varying values of ${\rm [Fe]}=-5$ to $-2$. Notations are the same as for Table~\ref{tab:yieldmass}.}
\vskip 0.1cm
\begin{tabular}{llllllrrrr}
\hline
 Model            &Sr           &Y           &Zr           &Ba             &Pb        &[Sr/Y]  &[Sr/Zr]   &[Sr/Ba]&$n_{\rm c}$\\
 \hline
$(25,-5,-1)$      & $2.16(-10)$ &$1.77(-11)$  &$1.31(-11)$  & $1.59(-16)$  &$0.00$        &$0.38$  &$0.88$    &$5.63$&$6.84$\\
$(25,-4,-1)$      & $1.96(\;\;-9)$  &$1.59(-10)$  &$1.15(-10)$  & $1.59(-15)$  &$0.00$        &$0.39$  &$0.89$    &$5.59$&$6.73$\\
$(25,-3,-1)$      & $1.57(\;\;-8)$  &$1.25(\;\;-9)$   &$8.79(-10)$  & $7.61(-15)$  &$0.00$        &$0.40$  &$0.91$    &$5.81$&$6.64$\\
$(25,-2,-1)$      & $9.50(\;\;-8)$  &$7.11(\;\;-9)$   &$4.76(\;\;-9)$   & $1.23(-14)$  &$0.00$        &$0.42$  &$0.95$    &$6.38$&$5.86$\\

\hline

\end{tabular}
\label{tab:yield25}
\end{table*}

In our models, when the \textsl{s}-process occurs, 
the primary $^{16}$O is the predominant isotope with a mass fraction of $\gtrsim 0.8$, which
makes $^{16}$O the dominant poison for stars with initial abundances of
${\rm [CNO]}\lesssim -1$. For stars with ${\rm [CNO]}>-1$, however, $^{25,26}$Mg become the dominant 
poisons because the neutron-capture cross section for $^{16}$O is much smaller than those for
$^{25,26}$Mg. In addition, the effectiveness of $^{16}$O as a neutron poison is greatly reduced at 
He-burning temperatures due to neutron regeneration through $^{17}$O$(\alpha,\mathrm{n})^{20}$Ne following
$^{16}$O$(\mathrm{n},\gamma){^{17}{\rm O}}$. Specifically, for the neutron density achieved during the \textsl{s}-process,
the rate of $^{16}$O$(\mathrm{n},\gamma){^{17}{\rm O}}$ is orders of magnitude slower than that of
$^{17}$O$(\alpha,\mathrm{n})^{20}$Ne. The latter is also a factor of $\kappa \sim 13$--15 higher 
than the rate of $^{17}$O$(\alpha,\gamma)^{21}$Ne at temperatures of $\sim (2.5$--$3) \times 10^8\,$K 
relevant for the \textsl{s}-process. As a result, the effective rate of neutron capture by $^{16}$O is
reduced by a factor of $\sim\kappa$.

The value of $\kappa \sim 13$--15 relevant for the \textsl{s}-process corresponds to the default rates \citep{cf88} 
for $^{17}$O$(\alpha,\mathrm{n})^{20}$Ne and $^{17}$O$(\alpha,\gamma)^{21}$Ne in our study. 
\citet{desc1993} gave a $\kappa$ three orders of magnitude 
larger. Such a large $\kappa$ would drastically increase the efficiency of 
the \textsl{s}-process in stars with initial abundances of ${\rm [CNO]}\lesssim -1$, for which $^{16}$O is the 
dominant neutron poison. Recent measurements by \citet{best2011,best2013}, however, gave a $\kappa$ within $\sim 10\%$ 
of our default value. We note that recent studies of the \textsl{s}-process in fast-rotating ``spinstars'' 
by \citet{frischk2016} and \citet{choplin2017} explored the effects of a lower rate for  
$^{17}$O$(\alpha,\gamma)^{21}$Ne citing unpublished measurements. To explore such a possibility as well,
we reduce the rate of $^{17}$O$(\alpha,\gamma)^{21}$Ne from its default value by a factor of 3 and 10, 
respectively, for $25 \,\mathrm{M}_\odot$ models with ${\rm [Fe]}=-3$ and varying [CNO]. 
As expected, the reduced rate has a strong effect on the \textsl{s}-process for models with ${\rm [CNO]}\lesssim -1$,
but has a rather small impact for ${\rm [CNO]}=0$, in which case $^{25,26}$Mg are the dominant neutron 
poisons (see Table~\ref{tab:o17agrate}). For example, when the rate is reduced by a
factor of 3, the Sr yield increases by a factor of 100, 13, and 1.3 for models with ${\rm [CNO]}=-2$, $-1$,
and 0, respectively. The reduced rate has an even stronger effect on the \textsl{s}-process flow beyond $^{88}$Sr:
when the rate is reduced by a factor of 10, the Zr yield increases $\sim5\times10^3$ 
and $\sim 10^2$ times for models with ${\rm [CNO]}=-2$ and $-1$, respectively.

\begin{table*}
\centering
\vspace{-0.5\baselineskip}
\caption{Post-CCSN yields (in $\,\mathrm{M}_\odot$) of Sr, Y, Zr, Ba, and Pb for
$25 \,\mathrm{M}_\odot$ models with ${\rm [Fe]}=-3$ and varying values of ${\rm [CNO]}=-2$ to 0
assuming different rates for $^{17}$O$(\alpha,\gamma)^{21}$Ne.
Notations are the same as for Table~\ref{tab:yieldmass}.}
\vskip 0.1cm
\begin{tabular}{lllllllrrrr}
\hline
 Model&$^{17}$O$(\alpha,\gamma)^{21}$Ne        
                    &Sr           &Y            &Zr           &Ba           &Pb         &[Sr/Y]  &[Sr/Zr]  &[Sr/Ba]&$n_{\rm c}$\\
 \hline
$(25,-3,-2)$&CF88   & $4.17(-13)$ &$9.01(-15)$  &$3.30(-15)$  & $0.00$          &$0.00$       &$0.96$   &$1.76$   & \nodata  &$1.06$\\
$(25,-3,-2)$&CF88/3&$4.18(-11)$&$1.21(-12)$&$3.18(-13)$& $3.14(-16)$  &$0.00$       &$0.83$&$1.78$&$5.29$&$2.06$\\
$(25,-3,-2)$&CF88/10&$8.39(-10)$&$4.24(-11)$&$1.58(-11)$&$2.88(-24)$&$0.00$&$0.59$&$1.38$&$13.96$&$3.55$\\
\hline
$(25,-3,-1)$&CF88   & $1.57(\;\;-8)$  &$1.25(\;\;-9)$    &$8.79(-10)$  & $7.61(-15)$ &$0.00$       &$0.40$  &$0.91$    &$5.81$&$6.64$\\
$(25,-3,-1)$&CF88/3 & $2.11(\;\;-7)$  &$2.31(\;\;-8)$   &$2.26(\;\;-8)$    & $1.04(-11)$ &$0.00$       &$0.26$  &$0.64$    &$3.80$&$11.44$\\
$(25,-3,-1)$&CF88/10& $6.95(\;\;-7)$  &$9.77(\;\;-8)$   &$1.22(\;\;-7)$    & $3.44(-10)$ &$1.89(-15)$&$0.15$ &$0.42$    &$2.80$&$16.35$\\
\hline
$(25,-3,\phantom{-}0)$&CF88   & $1.58(\;\;-6)$  &$3.13(\;\;-7)$   &$8.78(\;\;-7)$   & $3.10(\;\;-8)$  &$1.28(-11)$ &$0.00$    &$-0.09$    &$1.20$&$26.41$\\
$(25,-3,\phantom{-}0)$&CF88/3 & $2.09(\;\;-6)$  &$4.70(\;\;-7)$   &$1.54(\;\;-6)$   & $1.60(\;\;-7)$  &$2.49(-10)$ &$-0.06$&$-0.21 $   &$0.61$&$32.56$\\
$(25,-3,\phantom{-}0)$&CF88/10& $2.16(\;\;-6)$  &$4.93(\;\;-7)$   &$1.79(\;\;-6)$   & $2.86(\;\;-7)$  &$8.50(-10)$ &$-0.06$&$-0.26$    &$0.37$&$36.65$\\

\hline
\end{tabular}
\label{tab:o17agrate}
\end{table*}

The role of $^{16}$O as a poison also depends on the $^{16}$O$(\mathrm{n},\gamma){^{17}{\rm O}}$ rate.
The effect of this rate can be estimated by comparing the number of neutrons captured per available $^{56}$Fe seed, 
$n_{\rm c}=\Sigma_{A>56} [Y_{\!A}-Y_{\!A}(0)](A-56)/Y_{56}(0)$, where $Y_{\!A}(0)$ and $Y_{\!A}$ are the initial and final 
abundances, respectively, at mass number $A$.  Using a rate $\sim 10$ times lower than the 
updated value \citep{BAAL} adopted here, \citet{baraffe1992} found $n_{\rm c}\sim 12$ for a $30\,\mathrm{M}_\odot$ 
model with ${\rm [Fe]}\sim-2$ and ${\rm [CNO]}\sim -1.5$. Using a rate $\sim 100$ times lower than the updated value,
\citet{raiteri1992} found $n_{\rm c}\sim 8$ for a $25\,\mathrm{M}_\odot$ model 
with ${\rm [Fe]}\sim-2.3$ and ${\rm [CNO]}\sim -1.4$.  Ignoring neutron capture on $^{16}$O,
\citet{prantzos1990} found $n_{\rm c}\sim 8$--$18$ for $\sim 12$--$40\,\mathrm{M}_\odot$ models with ${\rm [Fe]}\sim-2$ and 
${\rm [CNO]}\sim -1.5$, in sharp contrast to $n_{\rm c}\sim 1$--$4$ for similar models in Table~\ref{tab:yieldmass}.

\subsection{Production of elements beyond {\rm Sr}}
The \textsl{s}-process flow slows down greatly when it encounters $^{88}$Sr with the magic neutron number $N=50$,
which usually marks the effective end point for the weak \textsl{s}-process. This feature can be clearly seen 
from Fig.~\ref{fig:cdep}, which shows steeply decreasing yields beyond Sr with negligible Ba 
production for models with ${\rm [CNO]}\lesssim-1$. 
For the most C-rich models with ${\rm [CNO]}=0$, however, substantial 
\textsl{s}-process flow proceeds beyond $^{88}$Sr for stars of $\gtrsim 20 \,\mathrm{M}_\odot$ with comparable yields of Sr 
and Zr (see Table~\ref{tab:yieldmass}). For stars of $\gtrsim 30 \,\mathrm{M}_\odot$, the \textsl{s}-process is even able to 
produce substantial amounts of Ba with [Sr/Ba] as low as $\sim 0.3$. As can be seen from 
Table~\ref{tab:o17agrate}, decreasing the $^{17}$O$(\alpha,\gamma)^{21}$Ne rate also results in considerable 
increase in the yields beyond Sr, especially for models with ${\rm [CNO]}\lesssim -1$.

\section{DISCUSSION AND CONCLUSIONS}
We have studied pre-CCSN nucleosynthesis in CEMP stars of $12$--$40\,\mathrm{M}_\odot$ with initial abundances of 
${\rm [Fe]}\leq-2$ and ${\rm [CNO]}=-2$ to $0$.  We find that the enhanced initial CNO abundances of such a star enable 
a weak \textsl{s}-process whose efficiency is determined by the [CNO] and the mass of the star and whose yields scale 
approximately linearly with the (Fe/H) of the star.  The \textsl{s}-process is especially efficient in stars of 
$\gtrsim 20 \,\mathrm{M}_\odot$ with ${\rm [CNO]}\gtrsim -1.5$, producing mainly elements up to Zr ($A\sim 90$) with 
${\rm [Sr/Zr]}\sim -0.3$ to 1.6. For the most C-rich (${\rm [CNO]}=0$) stars studied here, comparable amounts of Sr 
and Zr (${\rm [Sr/Zr]}\sim -0.2$) are produced in stars of $\gtrsim 30 \,\mathrm{M}_\odot$, along with substantial
amounts of Ba (${\rm [Sr/Ba]}\sim 0.3$--0.8). Whereas the default $^{17}$O$(\alpha,\gamma)^{21}$Ne rate of
\citet{cf88} used for our main results is in agreement with the published measurement of \cite{best2011},
reducing this rate can dramatically increase the weak \textsl{s}-process yields.
Unlike the main \textsl{s}-process in low to intermediate mass stars, the weak \textsl{s}-process in massive stars 
produces negligible amounts of Pb (see Tables~\ref{tab:yieldmass}--\ref{tab:o17agrate}).

As noted in the introduction, the \textsl{s}-process in massive CEMP stars is very similar to that in spinstars, which 
are fast-rotating massive VMP stars with normal initial CNO abundances \citep{pignatari2008,frischk2016}. The main 
difference between these two \textsl{s}-process models is in the production of the $^{22}$Ne that provides the neutron 
source. In CEMP stars the $^{22}$Ne is produced by burning the initial CNO, whereas in spinstars it is made by 
burning the primary $^{14}$N whose production is facilitated by rotation-induced mixing. 
It is difficult to assess the frequency of occurrences for spinstars in the early Galaxy.
In contrast, observations show that CEMP-no stars constitute $\sim 20\%$ of the low-mass VMP stars \citep{yong2013,placco2014}. 
As CEMP-no stars are thought to reflect the composition of the ISM polluted by the first or very early 
massive stars, formation of massive CEMP stars from the same ISM must also be relatively common.
In a recent compilation of 125 CEMP-no stars by \citet{yoonCEMPno1}, 24 ($\sim 19\%$) 
have ${\rm [C/H]}>-1.5$ with corrections for depletion during evolution \citep{placco2014}. Furthermore, 
12 ($\sim 10\%$) such stars have ${\rm [C/H]}>-1$ and 5 ($4\%$) have ${\rm [C/H]}>-0.5$. If this distribution of
[C/H] extends to massive CEMP stars, a significant fraction of them would have had an efficient \textsl{s}-process, and 
therefore, made important contributions of heavy elements to the early Galaxy. 

A recent study by \citet{hansenCEMPs} found that among the low-mass CEMP stars enhanced in heavy elements of the 
\textsl{s}-process origin, the so-called CEMP-\textsl{s} stars \citep{beers2005}, $\sim 10$--$30\%$ could be single stars.
The surface abundances of these stars would reflect the composition of their birth ISM instead of pollution by 
binary companions as for the rest of the CEMP-\textsl{s} stars. The origin of the heavy elements in single CEMP-\textsl{s} 
stars was investigated by \cite{pingest}. The C abundances of ${\rm [C/H]}\sim -0.5$ in some of
these stars \citep{spite2013} reinforce the indication from CEMP-no stars that some early ISM was highly 
enriched in C. Massive CEMP stars formed from such ISM would produce significant amounts of Ba by the 
\textsl{s}-process discussed here and contribute Ba and associated heavy elements to stars of the subsequent 
generation. 

The ejecta from a typical CCSN would be mixed with $\sim 10^3$--$10^4 \,\mathrm{M}_\odot$ of ISM.
Massive CEMP stars of $\gtrsim 25 \,\mathrm{M}_\odot$ with initial abundances of ${\rm [CNO]}=-1$ and ${\rm [Fe]}=-3$
would enrich the ISM with $\log \epsilon({\rm Sr})\sim -1.6$ to 0.6, 
which is in agreement with the range observed in many 
VMP stars with ${\rm [Fe/H]}\gtrsim-3$. Scaling the Sr yield with Fe, we obtain the enrichment by similar
CEMP stars but with ${\rm [Fe]}=-4$ to be $\log \epsilon({\rm Sr})\sim -2.6$ to $-0.4$, which is in agreement 
with the typical Sr abundances of VMP stars with $-4\lesssim {\rm [Fe/H]}\lesssim-3$ \citep{saga}. 
Likewise, CEMP stars of $\gtrsim 25 \,\mathrm{M}_\odot$ with initial abundances of ${\rm [CNO]}=0$ and ${\rm [Fe]}=-3$ 
would provide $-1.5 \lesssim \log \epsilon({\rm Ba})\lesssim 0.8$ along with 
$0.4 \lesssim \log \epsilon({\rm Sr})\lesssim 1.8$, and these results can be scaled to estimate the
enrichment by similar stars but with different [Fe]. In conclusion, the \textsl{s}-process in massive CEMP stars
has rather interesting implications for chemical evolution of the Early Galaxy.

 \section*{Acknowledgements}
 This work was supported in part by the NSFC [11533006 (SJTU), 11655002 (TDLI)], the US DOE [DE-FG02-87ER40328 (UM)], 
 the Science and Technology Commission of Shanghai Municipality [16DZ2260200 (TDLI)], and the ARC [FT120100363 (AH)].


\label{lastpage}
\end{document}